\documentclass[]{aa}
\usepackage{psfig}
\def\simlt{\lower.5ex\hbox{$\; \buildrel < \over \sim \;$}}
\def\simgt{\lower.5ex\hbox{$\; \buildrel > \over \sim \;$}}
\def\H2{\element[][][][2]{H}}

\begin{document}

   \msnr{Cg061}

   \title{An X-ray Binary Model for the Galactic Center Source IRS~13E}

   \author{R.F. Coker \and J.M. Pittard}

   \offprints{R.F. Coker}

   \institute{Department of Physics and Astronomy, \\ University of Leeds, 
              Leeds LS2 9JT  UK \\
              email: robc@ast.leeds.ac.uk,jmp@ast.leeds.ac.uk
             }

   \date{Received 2000 / Accepted ????}

   \thesaurus{01     
              (08.09.2 IRS 13E; 
               08.23.2; 
               13.25.5 
               )}

   \titlerunning{A Model for IRS 13E}
   \authorrunning{Coker \& Pittard}

   \maketitle

   \begin{abstract}

We present several models for IRS~13E, an infrared, mm and X-ray 
source in the Galactic Center. Our favored interpretation is that of
an early-type binary with strong colliding winds emission. This 
naturally explains the observed X-ray count rate and the strong IR 
emission lines, and has a distinct advantage over competing 
hypotheses based upon a single star or BH system. 
 
   \end{abstract}

\section{Introduction}

It is probable that Sgr A*, the compact, nonthermal radio source
at the Galactic Center (GC) is a 2--3$\times10^6 M_{\sun}$ black
hole (see, e.g., Ghez et al. 1998).  Deep in the potential well of
Sgr A* and pervading the central parsec of the Milky Way,
there exists a cluster of HeI and
late-type stars (see, e.g. Sellgren et al. 1990 and Genzel et al. 1996).
In order to understand the
evolution and dynamics of this unique stellar cluster,
it is necessary to
investigate individual stellar sources in detail.
IRS 13, identified as a Pa-$\alpha$, [FeIII], HeI, and HeII line
source (Stolovy et al. 1999; Lutz, Krabbe, \& Genzel 1993; 
Libonate et al. 1995; Krabbe et al. 1995),
is a compact HII region, dominated by the source IRS 13E.

Motivated by the {\it Chandra} results we speculate on the nature of
IRS~13 and conclude that it is most likely an early-type binary system.
This interpretation also supports the other observations.


\section{Source Identification}
\begin{figure}
       \psfig{figure=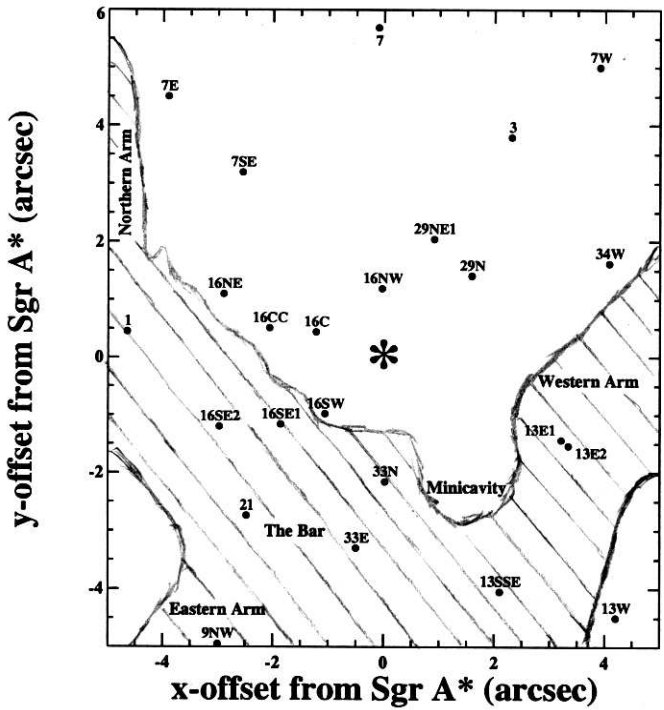,width=8.8cm}
      \caption[]{A schematic of the central half parsec of the Galaxy, showing the IRS
       sources which are discussed in this paper as well as other early-type stars
       in the region.  Shown in crosshatch are
       some of the ionized gas components of the central mini-spiral.  The location
       of Sgr A* is shown by an asterisk.  The locations of IRS 13E1 and IRS 13E2 are
       from Ott et al. (1999) and assume that IRS 7 is offset from Sgr A* by -0.1 and -5.69
       arcseconds in x and y, respectively.}
      \label{fig:map}
\end{figure}

Due to the $\sim 30$ magnitudes of visual extinction towards the GC 
(Wade et al 1987; Blum, Sellgren, \& Depoy, 1996), the only observations of IRS 13 are at radio, 
infrared and X-ray wavelengths. At $\lambda=2$ cm, the IRS 13 complex 
is one of the brightest sources in the central $10''$ of the Galaxy 
(Yusef-Zadeh, Roberts, \& Biretta 1997). Observations at 7mm and 13mm, 
where interstellar scatter-broadening is less significant, resolve 
IRS~13 into two sources. Infrared K-band photometry also
resolves IRS 13E into two sources (IRS 13E1 and IRS 13E2),
the former being identified with a Wolf-Rayet star 
of type WN9-10 (Najarro et al. 1997, hereafter N97). The two sources 
are separated by $0.16'' \sim 1000$ AU in projection (assuming a 
GC distance of 8 kpc) with positional uncertainties of $\sim 0.1''$ 
(Ott, Eckart \& Genzel 1999).  This compares to a separation of 
$\sim 500$~AU for the mm sources although the positional uncertainties 
are consistent with these mm sources\footnote{Zhao \& Goss (1999) 
label these source as IRS 13E and IRS 13W;
here, we will identify them as IRS 13E1 and IRS 13E2, respectively.}
being the same as the infrared sources identified by Ott, Eckart 
\& Genzel (1999).  Fig.~\ref{fig:map} presents a schematic of the region, 
showing the mini-spiral and various wind sources in the central 10
arcseconds.

A recent {\it Chandra} observation of the GC (Baganoff et al. 1999) 
shows a compact spherical X-ray source apparently coincident with 
IRS 13, located $\sim 3''$ ESE from Sgr A*. With $\sim 100$ counts 
after a 50 ksec observation, it is nearly as bright as Sgr A*. The 
only other point source in the central $\sim 10''$ is a more 
asymmetrical object, with no obvious radio or infrared counterpart, 
that is located $\sim 9''$ NNE of Sgr A*. 

\section {Radio, sub-mm and IR observations}

(Zhao \& Goss 1999) determined a spectral index of $0.9\pm0.2$ for 
IRS~13E1, similar to that expected from a fully ionized stellar 
wind (Wright \& Barlow 1975; Nugis, Crowther, \& Willis 1998).
Conversely IRS 13E2 has a flat mm spectrum which may be nebular 
(Whiteoak 1994). Such a mixture of compact and extended emission is 
similar to the Luminous Blue Variable (LBV) $\eta$ Car (see, e.g., 
Cox et al 1995), which may have a binary companion (Damineli et al. 2000).

A major feature of IRS~13 is the strength of some of its IR emission
lines. IRS 13 has stronger $2.19 \mu$m HeII (N97) and Pa-$\alpha$ 
(Stolovy et al. 1999) lines than its nearby IRS 16 brethren
and is one of the strongest Br-$\gamma$ sources in the central 
parsec of the Galaxy (Libonate et al. 1995; Roberts, Yusef-Zadeh, 
\& Goss 1996; Herbst et al. 1993). However, IRS~13 has a rather weak
H$92\alpha$ line; Roberts et al (1996) thus suggest that the H$92\alpha$ source 
is hotter, denser and more compact than the Br$\gamma$ source 
(Roberts, Yusef-Zadeh, \& Goss 1996). Although its [FeII] emission is
weak (Stolovy et al. 1999), IRS 13 is located at the local peak 
of [FeIII] emission (Lutz, Krabbe, \& Genzel 1993), suggesting a 
local ionizing source. It should be noted, however, that
for some of the above observations, line blending (e.g. HeI(4-3) 
with Pa-$\alpha$ and HI(12-4) with the 1.644 $\mu$m [FeII] line)
can be a significant problem.

From observations of the K-band HeI and Br$\gamma$ emission lines, N97 
constructed non-LTE atmospheric models of IRS~13E1 and determined the 
parameters given in Table~\ref{tab:naj}. At first glance this mass loss 
rate is substantially more than the canonical upper limit of 
$\sim 1\times10^{-4}~\mathrm{M}_{\sun}$ yr$^{-1}$ for WR stars (Nugis 1999).
However, a number of arguments exist against such a high estimate. First
N97 did not include wind clumping or line-blanketing and as a result probably
over estimated the mass loss rate by at least a factor of 2 (Morris 
et al. 2000). Second, assuming the parameters in Table~\ref{tab:naj}, 
the classical stellar wind theory of Wright \& Barlow (1975)
predicts a 13 mm flux of $\sim 30$ mJy, considerably higher than the observed
$\sim 10$ mJy for the combined IRS 13E complex, again suggesting
an overestimate of $\dot\mathrm{M}/v_\infty$ by at least
a factor of 2. Third, if the other GC wind sources have similarly
overestimated mass loss rates, hydrodynamical accretion simulations 
(e.g. Coker \& Melia 1997), which predict a large mass accretion rate 
onto Sgr A*, are brought more into agreement with other accretion 
models (Coker \& Melia, 2000; Narayan et al. 1998), which require a much
lower accretion rate. For completeness, we also note that if the
metallicity of IRS 13E is more than twice solar (see, e.g., 
Shields \& Ferland 1994; Simpson et al. 1995), the maximum possible 
WR mass loss rate could be higher (Maeder \& Maynet 1987).

If we assume the metallicity of the GC is twice solar then
based on the estimated bolometric luminosity of IRS 13E1 
the present mass and zero-age main-sequence (ZAMS) mass
of IRS 13E1 are $\sim 60~\mathrm{M}_{\sun}$ and more than 
120~$\mathrm{M}_{\sun}$, respectively (Schaller et al 1992). However, 
updated evolutionary models (e.g., Mowlavi et al 1998) show that a 
high metallicity star with ZAMS mass of $\simgt 60~\mathrm{M}_{\sun}$ 
will shed so much mass while on the main sequence that it would 
never reach the WR stage. The canonical assumption that the metallicity
of the GC is twice solar, however, is based on gas phase abundances.
Carr, Sellgren, \& Balachandran (2000) further argue that GC stellar
abundances may in turn be overestimated due to higher stellar 
rotation rates. For example, high-resolution near-infrared spectra 
of IRS 7, a red supergiant located in the central parsec, show 
solar metallicity.  

\begin{table}
  \caption{Computed Model for IRS 13E1 (from N97)}
  \label{tab:naj}
  \begin{tabular}{ll}
    \hline
 & \\[-7pt]
 $\dot \mathrm{M}~(10^{-4} \mathrm{M}_{\sun}$yr$^{-1})$&7.9             \\
 v$_\infty$ ($10^3$ km s$^{-1})$              &1.0            \\
 L$_*~(10^6 \mathrm{L}_{\sun})$               &2.3             \\
 T$_\mathrm{eff}$ ($10^4$ K)                  &2.9             \\
 He/H                                         &$>500$           \\[3pt]
    \hline
  \end{tabular}
\end{table}

\section{What is IRS 13E?}

It is certain that IRS 13E is a luminous object embedded in a hot, 
dense medium, but more than this is not clear. However, it is possible
to explore a few options which best fit the observations. First,
the estimated luminosity and mass loss rate of IRS~13E1 are too large 
for a single evolved WR star; they are more like that of an LBV.
In addition Libonate et al. (1995) noted that the K-band spectrum of 
IRS~13 resembles that of the LBV, P Cyg. Similarly, the FWHM of the 
Br-$\gamma$ line ($\sim 350$ km sec$^{-1}$) as well as the extent 
of the emission region ($\simlt 0.12$ pc) (Herbst et al. 1993) is 
consistent with the idea that IRS 13E has recently undergone 
an LBV-like mass ejection, and also ties in well with the radio and
sub-mm observations. However, in spite of this agreement its 
temperature and wind velocity are larger than a typical LBV. As 
has been suggested for IRS 16NE, IRS 16C, and IRS 16SW
(N97), it could therefore be a transition object just coming out of the
LBV phase.  This would be consistent with the $v$ and $\dot\mathrm{M}$ 
profiles of Garcia-Segura, Mac Low \& Langer (1996).
Humphreys \& Davidson (1994) point out that a WN9/Ofpe star appears 
very much like an LBV at minimum brightness and thus the
distinction between the two types is often blurred.
However, the large He/H of IRS~13E1, compared to a He/H of $\sim 1$ 
for the IRS 16 sources (N97), makes an LBV/WNL determination problematic.
Observational and theoretical counterarguments include the fact that 
the identity of WR 122, the calibrating WNL source used by N97, has been
called into doubt (Crowther \& Smith 1999) and 
recent work (e.g., Langer et al. 1999) suggests
mixing in massive stars is more efficient than previously thought, resulting
in a larger He/H at the start of the WNL phase. Hence an LBV/WNL
classification is certainly within reason (although see below).

Second, we note that there are $\simgt 27$ massive HeI stars in 
the GC (Blum, Ram{\'i}rez, \& Sellgren 1999). Given the observed
frequency, $f$, of WR binary systems ($12\% \simlt f \simlt 50\%$; 
van der Hucht et al. 1981), it is likely that some of the GC
HeI stars are binaries as well. In fact, IRS~16SW is thought to be 
an eclipsing binary with a period of $\sim 10$ days (Ott, Eckart 
\& Genzel 1999). Thus it is possible that IRS~13E is also a binary
system, containing for example a WN10 primary with a ZAMS mass 
of $\sim 100~\mathrm{M}_{\sun}$ and a somewhat less massive companion 
that is either an O star or another WR star, or, possibly,
a massive compact object.

\section{Consequences}

Of the single and binary scenarios, the latter is preferred. As 
previously mentioned, the model of N97 underpredicts the K-band 
HeII emission for IRS~13E1 by a factor of $\sim 3$; the colliding 
winds of a binary will produce more He$^+$ (Marchenko et al. 1997),
potentially explaining this deficiency.  Also, the ionizing flux 
from a massive O star would explain why IRS~13E stands out in 
Pa-$\alpha$ and [FeIII].

An early-type binary system will also have strong shocks as a result
of colliding stellar winds.  
The X-ray luminosity of a binary system will be brighter than
that from a solitary star so that in order of increasing X-ray luminosity
one qualitatively has (with everything else equal) WR, O, O/O, 
WR/O, WR/WR.  However, this is modified by the binary separation: 
if the binaries are too close, absorption will suppress the observed
X-ray emission but if they are too far apart the shocks are largely 
adiabatic and do not produce significant
additional X-ray luminosity (Pittard \& Stevens 1997).

The various WR sources of IRS 16 (including IRS 16SW)
do not appear as point sources in the {\sl Chandra} observations.  
A solitary WR or even a WR with a B or later companion may not
be visible with {\sl Chandra} due to the large column density between 
here and the GC. Using {\sc XSPEC}\footnote{Distributed and maintained
by HEASARC} and a Raymond-Smith thermal plasma model (which assumes 
optically thin X-ray line and continuum emission; Raymond \& Smith 1977) 
in ionization equilibrium one can simulate 
the {\it Chandra} spectrum of a solitary O-star placed at the GC. 
Assuming an ISM-corrected X-ray luminosity (0.5-10.0~keV) 
$\mathrm{L_x} \sim 0.25~\mathrm{L}_{\sun}$ (from $L_{x}/L_{bol} 
\sim 10^{-7}$, see e.g. Waldron, et al. 1998), a characteristic 
temperature $\mathrm{kT} \sim 0.5$~keV (representative of typical
solitary O-stars, see Chlebowski, et al. 1989), 
an intervening column density N$_\mathrm{H} 
\sim 5 \times 10^{22}$ cm$^{-2}$ (typical for the GC, see e.g., 
Zylka et al. 1995), and solar abundances throughout, 
a 50 ksec {\it Chandra} observation, spanning 
the 0.5-10 keV band, would detect $\simlt 3$ photons, and the O-star
would not stand out above the background. Thus the IRS 16 sources 
blend into the diffuse background seen in the central $\sim 5''$ of
the image in Baganoff et al. (1999). Note that this hints that either 
the unseen IRS 16SW companion does not have a significant stellar 
wind and thus is of type B or later (and therefore implies
a lower system mass than estimated by Ott et al. 1999), or that the
circumstellar absorption and/or binary separation were unfavorable.

\begin{figure}
       \psfig{figure=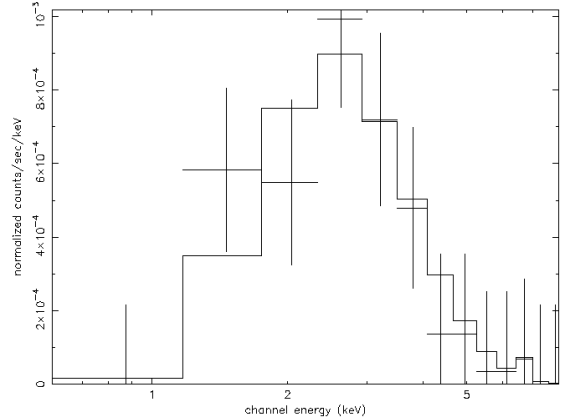,width=8.8cm}
      \caption[]{Plot of a model WR/O spectrum detected by {\sl Chandra} with a 50 ksec observation,
       assuming an intrinsic (0.5-10.0~keV) X-ray luminosity of $0.5~\mathrm{L}_{\sun}$ and a 
       characteristic temperature of 1.5 keV (both of which are typical for colliding wind binaries, 
       see ,{\sl e.g.}, Maeda, et al. 1999, Stevens, et al. 1996), and 
       a column density of $5.0 \times 10^{22}$~cm$^{-2}$.
       {\sl Chandra} would see a total of $\sim$ 100 photons from this hypothetical binary system.}
       \label{fig:WRO}
\end{figure}

\begin{figure}
      \psfig{figure=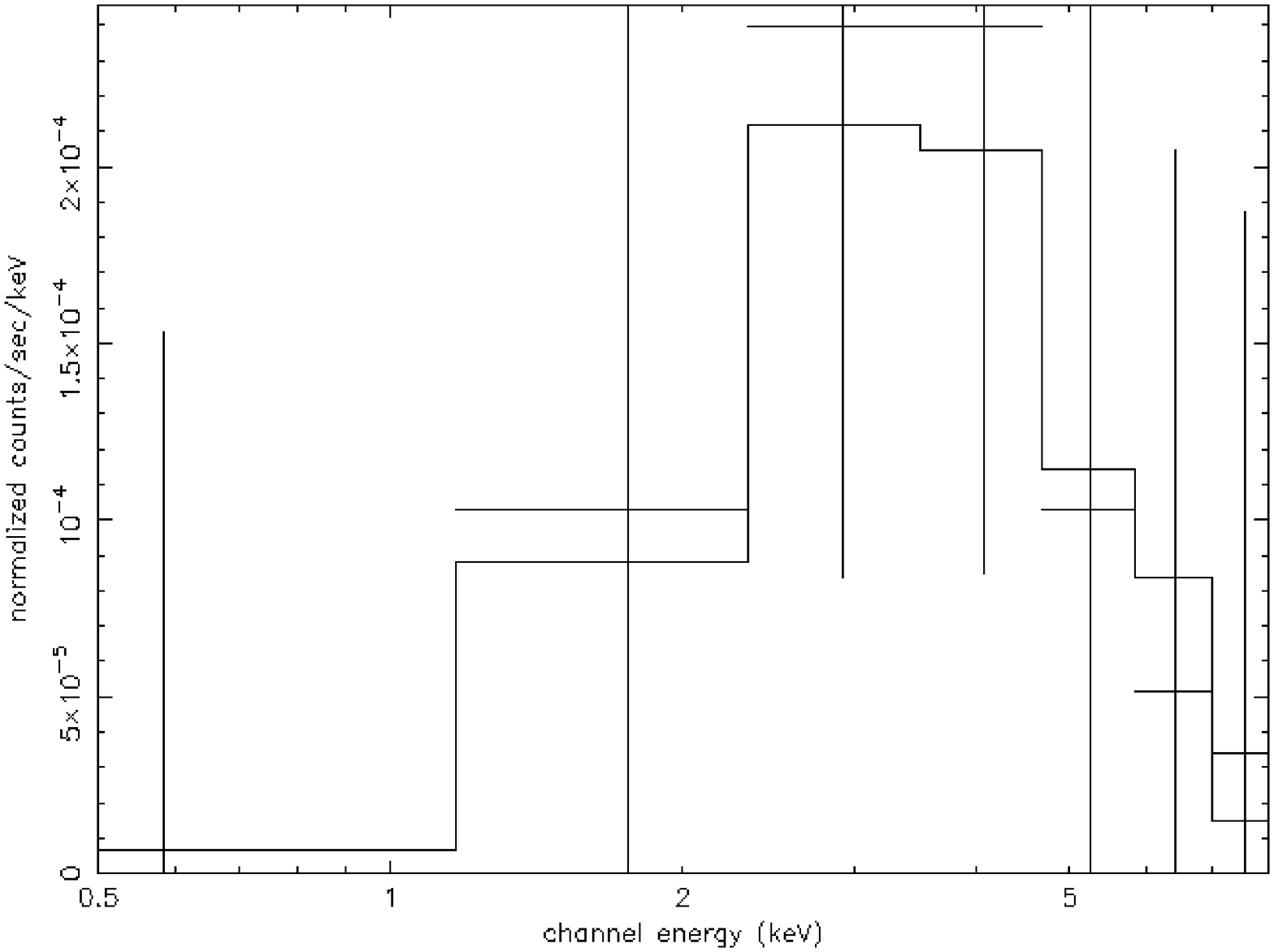,width=8.8cm}
      \caption[]{Plot of a model WR/BH spectrum detected by {\sl Chandra} with a 50 ksec observation,
       assuming an intrinsic luminosity of $0.1~\mathrm{L}_{\sun}$, a characteristic temperature of 5.0 keV, and 
       a column of $5.0 \times 10^{22}$~cm$^{-2}$.
       {\sl Chandra} would see a total of about 45 photons from this hypothetical source.}
      \label{fig:WRBH}
\end{figure}

In contrast, Fig.~\ref{fig:WRO} shows a theoretical {\it Chandra} 
spectrum for a hypothetical $0.5~\mathrm{L}_{\sun}$ GC X-ray source 
similar to the WR/O binary $\gamma^2$ Vel, which has a 78.5 day period. 
The number of photons detected by a 50 ksec observation of such a 
source is $\sim 100$, consistent with the supposed IRS 13 source seen
by Baganoff (1999). Note that the mm observations of IRS~13E2 probably 
do not correspond to the WR's companion but rather to the extended 
emission due to the the ejected nebula from the WR's LBV phase. 
Thus the binary separation is not constrained to match that of the
resolved mm sources.

Another possible interpretation of IRS~13 which would explain the 
{\it Chandra} observation follows from the suggestion by Gerhard (2000) 
that the GC HeI cluster is the remains of a disrupted cluster that 
formed tens of parsecs from Sgr A*. If true, a massive binary system 
may not have survived the infall. This raises the possibility that 
IRS~13E2 hides a compact object, such as a 
$\sim 10~\mathrm{M}_{\sun}$ black-hole, which is accreting the wind 
from IRS~13E1. That is, IRS~13E may be an extended X-ray
system. This is particularly applicable in GC since dynamical friction
over the lifetime of the Galaxy is likely to result in a high density 
of $\sim 10~\mathrm{M}_{\sun}$ black-holes in the central parsec 
(see, e.g., Morris 1993). One can estimate the X-ray luminosity of such 
a system using 
\begin{equation}\label{eq:xraybin}
L_x = \epsilon \dot\mathrm{M} c^2 \left( {{G M}\over{2 D v^2}}\right)^2 \;,
\end{equation}
where $\epsilon$ is the accretion efficiency (taken as 0.1), $M$ is 
the mass of the black-hole, $v$ is the relative velocity of the 
black-hole to the wind, and $D$ is the distance between the WR
star and the black-hole.  If $M = 10~\mathrm{M}_{\sun}$, $v = 1000$ 
km sec$^{-1}$, and $D = 500$ AU, then $L_x \sim 0.1~\mathrm{L}_{\sun}$. 
The spectrum, as shown in Fig.~\ref{fig:WRBH}, would be shifted to
slightly higher energy, the flux would be less attenuated
by the high column and thus the $\sim 45$ detected photons would be 
consistent with the IRS 13 detection of Baganoff (1999). Note that, if 
the $\sim 10$ mas yr$^{-1}$ relative mm proper motions 
(Zhao \& Goss 1999) are any indication, the system may be unbound, 
making IRS 13E a unique transient X-ray source.

In summary, whilst a single massive star of unusually high
X-ray luminosity cannot be discounted, we believe a colliding winds
binary system best fits the various observations. A BH system also
has some difficulties with reconciling the resolved mm sources (unless
the second source is unassociated). 
Long term monitoring at mm wavelengths as well as a long-integration 
{\it Chandra} observation would help determine precisely what type of 
object lurks in IRS 13E.  

\begin{acknowledgements}
This work was supported by PPARC and has made use of
NASA's Astrophysics Data System Abstract Service.  We thank Fred Baganoff 
for early discussions of his work and Angela Cotera for an inciteful 
review of the manuscript.
\end{acknowledgements}


\begin{thebibliography}{}

\bibitem[Baganoff, et al. (1999)]{1999AAS...195.6201B} Baganoff, F., and 14
colleagues 1999, American Astronomical Society Meeting, 195, 6201             

\bibitem[Blum, Sellgren and Depoy (1996)]{1996ApJ...470..864B} Blum, R. D., 
Sellgren, K. and Depoy, D. L. 1996, ApJ, 470, 864 

\bibitem[Blum, Ram{\'i}rez and Sellgren (1999)]{1999cpg..conf..291B} Blum,
R. D., Ram{\'i}rez, S. V. and Sellgren, K. 1999, ASP Conf. Ser. 186: The
Central Parsecs of the Galaxy, 291            

\bibitem[Carr, Sellgren and Balachandran (2000)]{2000ApJ...530..307C} Carr,
J. S., Sellgren, K. and Balachandran, S. C. 2000, ApJ, 530, 307                 
\bibitem[Chlebowski, et al. (1989)]{1989ApJ...341...427} Chlebowski, T.,
Harnden, F.R. Jr. and Sciortino, S. 1989, ApJ, 341, 427

\bibitem[Coker and Melia (2000)]{2000ApJ...534..723C} Coker, R. F. and
Melia, F. 2000, ApJ, 534, 723                              

\bibitem[Coker and Melia (1997)]{1997ApJ...488L.149C} Coker, R. F. and
Melia, F. 1997, ApJ, 488, L149                   

\bibitem[Cotera, et al. (1999)]{1999cpg..conf..240C} Cotera, A., Morris,
M., Ghez, A. M., Becklin, E. E., Tanner, A. M., Werner, M. W. and Stolovy,
S. R. 1999, ASP Conf. Ser. 186: The Central Parsecs of the Galaxy, 240             

\bibitem[Cox, et al. (1995)]{1995A&A...297..168C} Cox, P., Mezger, P. G.,
Sievers, A., Najarro, F., Bronfman, L., Kreysa, E. and Haslam, G. 1995,
A\&A, 297, 168

\bibitem[Crowther and Smith (1999)]{1999MNRAS.308...82C} Crowther, P. A.
and Smith, L. J. 1999, MNRAS, 308, 82                     

\bibitem[Damineli, et al. (2000)]{2000ApJ...528..L101} Damineli, A.,
Kaufer, A., Wolf, B., Stahl, O., Lopes, D. F. and de Ara\'{u}jo, F. X. 2000, ApJ,
528, L101                        

\bibitem[Garcia-Segura, Mac Low and Langer (1996)]{1996A&A...305..229G}
Garcia-Segura, G., Mac Low, M.-M. and Langer, N. 1996, A\&A, 305, 229            

\bibitem[Genzel, et al. (1996)]{1996ApJ...472..153G} Genzel, R., Thatte, 
N., Krabbe, A., Kroker, H. and Tacconi-Garman, L. E. 1996, ApJ, 472, 153 

\bibitem[Gerhard (2000)]{2000APJ...submitted}
Gerhard, O., ApJ, submitted, astro-ph/0005096        

\bibitem[Gezari and Yusef-Zadeh (1990)]{1990awia.conf..214G} Gezari, D. and
Yusef-Zadeh, F. 1990, ASP Conf. Ser. 14: Astrophysics with Infrared Arrays,
214             

\bibitem[Ghez, Klein, Morris and Becklin (1998)]{1998ApJ...509..678G} Ghez,
A. M., Klein, B. L., Morris, M. and Becklin, E. E. 1998, ApJ, 509, 678         

\bibitem[Herbst, Beckwith, Forrest and Pipher (1993)]{1993AJ....105..956H}
Herbst, T. M., Beckwith, S. V. W., Forrest, W. J. and Pipher, J. L. 1993,
AJ, 105, 956

\bibitem[Krabbe, et al. (1995)]{1995ApJ...447L..95K} Krabbe, A., and 13
colleagues 1995, ApJ, 447, L95        

\bibitem[Langer, Heger, Wellstein and Herwig (1999)]{1999A&A...346L..37L}
Langer, N., Heger, A., Wellstein, S. and Herwig, F. 1999, A\&A, 346, L37         

\bibitem[Libonate, Pipher, Forrest and Ashby (1995)]{1995ApJ...439..202L}
Libonate, S., Pipher, J. L., Forrest, W. J. and Ashby, M. L. N. 1995, ApJ,
439, 202
                   
\bibitem[Lutz, Krabbe and Genzel (1993)]{1993ApJ...418..244L} Lutz, D.,
Krabbe, A. and Genzel, R. 1993, ApJ, 418, 244             

\bibitem[Maeda, et al. (1999)]{1999ApJ...510...967} Maeda, Y., Koyama, K.,
Yokogawa, K. and Skinner, S. 1999, ApJ, 510, 967

\bibitem[Marchenko, et al. (1997)]{1997ApJ...485..826M} Marchenko, S. V.,
Moffat, A. F. J., Eenens, P. R. J., Cardona, O., Echevarria, J. and
Hervieux, Y. 1997, ApJ, 485, 826   

\bibitem[Morris, et al. (2000)]{2000A&A...353..624M} Morris, P. W., van der
Hucht, K. A., Crowther, P. A., Hillier, D. J., Dessart, L., Williams, P. M.
and Willis, A. J. 2000, A\&A, 353, 624

\bibitem[Mowlavi, et al. (1998)]{1998A&AS..128..471M} Mowlavi, N.,
Schaerer, D., Meynet, G., Bernasconi, P. A., Charbonnel, C. and Maeder, A.
1998, A\&AS, 128, 471       
                                                 
\bibitem[Najarro, et al. (1997)]{1997A&A...325..700N} Najarro, F., Krabbe,
A., Genzel, R., Lutz, D., Kudritzki, R. P. and Hillier, D. J. 1997, A\&A,
325, 700 (N97)     

\bibitem[Narayan, et al. (1998)]{1998ApJ...492..554N} Narayan, R.,
Mahadevan, R., Grindlay, J. E., Popham, R. G. and Gammie, C. 1998, ApJ,
492, 554           

\bibitem[Nugis, Crowther and Willis (1998)]{1998A&A...333..956N} Nugis, T.,
Crowther, P. A. and Willis, A. J. 1998, A\&A, 333, 956

\bibitem[Nugis (1999)]{1999IAUS..193...84N} Nugis, T. 1999, IAU Symp. 193:
Wolf-Rayet Phenomena in Massive Stars and Starburst Galaxies, 193, 84        

\bibitem[Ott, Eckart and Genzel (1999)]{1999ApJ...523..248O} Ott, T.,
Eckart, A. and Genzel, R. 1999, ApJ, 523, 248      

\bibitem[Pittard and Stevens (1997)]{1997MNRAS.292..298P} Pittard, J. M.
and Stevens, I. R. 1997, MNRAS, 292, 298                                 

\bibitem[Raymond and Smith (1977)]{1977ApJS...35..419} Raymond, J. C. and
Smith, B. W. 1977, ApJS, 35, 419      

\bibitem[Roberts, Yusef-Zadeh and Goss (1996)]{1996ApJ...459..627R}
Roberts, D. A., Yusef-Zadeh, F. and Goss, W. M. 1996, ApJ, 459, 627    

\bibitem[Schaller, Schaerer, Meynet and Maeder (1992)]{1992A&AS...96..269S}
Schaller, G., Schaerer, D., Meynet, G. and Maeder, A. 1992, A\&AS, 96, 269
                                                                                     
\bibitem[Sellgren, McGinn, Becklin and Hall (1990)]{1990ApJ...359..112S} 
Sellgren, K., McGinn, M. T., Becklin, E. E. and Hall, D. N. 1990, ApJ, 
359, 112 

\bibitem[Shields and Ferland (1994)]{1994ApJ...430..236S} Shields, J. C.
and Ferland, G. J. 1994, ApJ, 430, 236

\bibitem[Simpson, et al. (1995)]{1995ApJ...444..721S} Simpson, J. P.,
Colgan, S. W. J., Rubin, R. H., Erickson, E. F. and Haas, M. R. 1995, ApJ,
444, 721              

\bibitem[Stolovy, et al. (1999)]{1999cpg..conf...39S} Stolovy, S. R., 
McCarthy, D. W., Melia, F., Rieke, G., Rieke, M. J. and Yusef-Zadeh, F. 
1999, ASP Conf. Ser. 186: The Central Parsecs of the Galaxy, 39 

\bibitem[van der Hucht, et al. (1981)]{1986SSR...28..227} van der Hucht, K. A.,
Conti, P. S., Lundstr\"{o}m, I. and Stenholm, B. 1981, SSR, 28, 227            

\bibitem[Wade, et al. (1987)]{1987ApJ...320..570W} Wade, R., Geballe, T.
R., Krisciunas, K., Gatley, I. and Bird, M. C. 1987, ApJ, 320, 570          

\bibitem[Waldron, et al. (1998)]{1998ApJS...118...217} Waldron, W. L., 
Corcoran, M. F., Drake, S. A. and  Smale, A. P., 1998, ApJS, 118, 217

\bibitem[Wright and Barlow (1975)]{1975MNRAS.170...41W} Wright, A. E. and
Barlow, M. J. 1975, MNRAS, 170, 41
                                               
\bibitem[Yusef-Zadeh, Roberts and Biretta (1997)]{1997AAS...191.9704Y}
Yusef-Zadeh, F., Roberts, D. A. and Biretta, J. 1997, American Astronomical
Society Meeting, 191, 9704         

\bibitem[Zhao and Goss (1999)]{1999cpg..conf..224Z} Zhao, J.-H. and Goss,
W. M. 1999, ASP Conf. Ser. 186: The Central Parsecs of the Galaxy, 224             

\bibitem[Zylka, et al. (1995)]{1995A&A...297...83Z} Zylka, R., Mezger, P.
G., Ward-Thompson, D., Duschl, W. J. and Lesch, H. 1995, A\&A, 297, 83       

\end{thebibliography}
\end{document}